\documentclass[aps,pre,
preprint,groupedaddress,superscriptaddress,
showpacs]{revtex4-1}

\usepackage{amsmath}
\usepackage{amssymb}
\usepackage{euscript}
\usepackage{bm}
\usepackage{mathtools}

\usepackage{graphicx}
\usepackage{epstopdf}
\usepackage[usenames,dvipsnames]{xcolor}
\usepackage{caption}
\usepackage{subcaption}
\usepackage{color}
\usepackage{ulem}

\newcommand{\vea}{\varepsilon_a}

\begin{document}

\title{Energy surface and minimum energy paths for Fr\'{e}edericksz transitions in bistable cholesteric liquid crystals}

\author{A.V.~Ivanov} \email{alekcey92@inbox.ru}
\affiliation{ Saint Petersburg State University, 199034, Saint Petersburg, Russia}
\affiliation{Science Institute of the University of Iceland, VR-III, 107 Reykjavik, Iceland}

\author{P.F.~Bessarab}
\affiliation{ Saint Petersburg State University, 199034, Saint Petersburg, Russia}
\affiliation{Department of Materials and Nanophysics, Royal Institute of Technology (KTH), SE-16440 Kista, Sweden}

\author{E.V.~Aksenova}
\affiliation{ Saint Petersburg State University, 199034, Saint Petersburg, Russia}

\author{\fbox{V.P.~Romanov}}
\affiliation{ Saint Petersburg State University, 199034, Saint Petersburg, Russia}

\author{V.M.~Uzdin}
\affiliation{ Saint Petersburg State University, 199034, Saint Petersburg, Russia}
\affiliation{Saint Petersburg National Research University of Information Technologies, Mechanics and Optics, 197101 Saint Petersburg, Russia}

\begin{abstract}
The multidimensional energy surface of a cholesteric liquid crystal in a planar cell is investigated as a function of spherical coordinates determining the director orientation. Minima on the energy surface correspond to the stable states with particular director distribution. External electric and magnetic fields deform the energy surface and positions of minima. It can lead to the transitions between states, known as the Fr\'{e}edericksz effect. Transitions can be continuous or discontinuous depending on parameters of the liquid crystal which determine an energy surface. In a case of discontinuous transition when a barrier between stable states is comparable with the thermal energy, the activation transitions may occur and it leads to the modification of characteristics of the Fr\'{e}edericksz effect with temperature without explicit temperature dependencies of liquid crystal parameters. Minimum energy path between stable states on the energy surface for the Fr\'{e}edericksz transition is found using geodesic nudged elastic band method. Knowledge of this path, which has maximal statistical weight among all other paths, gives the information about a barrier between stable states and configuration of director orientation during the transition. It also allows one to estimate the stability of states with respect to the thermal fluctuations and their lifetime when the system is close to the Fr\'{e}edericksz transition.
\end{abstract}

\date{\today}



\maketitle

\section{Introduction}
\label{sec1}

Cholesteric liquid crystals (ChLC) have attracted much attention because of their vast applications in optical systems, including display devices~\cite{yang_94a,lu_97,wu_01}, light shutters and optical filters~\cite{huang_03,mitov_04,mitov_12}, lasers~\cite{kopp_98,coles_10,dolgaleva_08} and others. Under certain conditions, ChLCs are characterized by co-existence of several stable states with distinct optical properties. Multistability of ChLC is particularly important for the applications in energy-efficient optical display systems such as bistable reflective displays~\cite{yang_94a,lu_97} where visual information is maintained at a zero power consumption and the only energy loss is associated with the refresh of displayed data.

Co-existence of the planar (P) state, where the director forms a perfect helix confined in the liquid crystal cell, and the focal conic (FC) state, where the helix pitch becomes irregular and the helix axis acquires a component parallel to the cell surface, was first observed more than 40 years ago~\cite{greubel_73}. Since then, several methods have been developed to obtain bistability in ChLC, including admixture of dispersed polymer~\cite{john_95}, special treatment of the cell surfaces~\cite{yang_94b} as well as application of external fields. Bistability of P and FC states can be achieved even at zero external field~\cite{yang_94a}, which is particularly important for the applications. 

Theoretical description of transitions between the stable states in ChLC is an important problem in fundamental studies of liquid crystals (LC) and is of critical importance in the design of optical liquid crystal displays, where efficient switching between the optical states is needed for recording the visual information. Several schemes have been proposed to induce transitions between stable states in ChLC, involving application of the external field pulses~\cite{yang_94a,crawford_96,oh_14,kim_10,lu_14} and pressure~\cite{kim_11}. Thermal fluctuations can induce spontaneous transitions and, therefore, affect the stability of optic states in ChLC. The preparation of a ChLC system in a particular state can be destroyed by thermally activated transitions to other available states. Typically, P and FC structures are very stable against thermal fluctuations due to the large energy barrier separating the states~\cite{crawford_96,lu_14,wang_11}. However, energy barriers can be tuned by external fields driving the system to the regime where spontaneous thermally activated transitions can not be neglected. Thermally assisted switching between P and FC states has already been proposed for recording visual data in ChLC devices~\cite{fuh_10}. 

In this article, we study transitions between stable states in ChLC by analyzing the multidimensional energy surface of the system defined by the Oseen-Frank model~\cite{degennes_93}. The minima on the energy surface correspond to stable states, while minimum energy paths (MEPs) between them define the mechanism of transitions. A MEP represents the path that lies lowermost on the energy surface and a maximum along the MEP corresponds to a saddle point (SP), which gives the energy barrier. We apply geodesic nudged elastic band (GNEB) method~\cite{bessarab_2015} to calculate MEPs of transitions between stable states in ChLC. We study how external electric and magnetic fields as well as boundary conditions affect energy barriers. Analysis of the energy barriers is needed for the quantitative assessment of the effect of thermally activated transitions within the rate theory~\cite{coffey_01}. In particular, we show that thermal activation needs to be taken into account when assessing the stability of states in ChLC with respect to an applied field, contributing to the temperature dependence of the transition field.
Based on the Oseen-Frank model, we also develop a reduced description of the ChLC providing a two-dimensional representation of the energy surface of ChLC, where the minima, SPs and MEPs can be visualized giving an insight into the transition mechanism and effect of the applied field on the transition path.

This article is organized as follows. 
In Sec.~\ref{sec2}, a multidimensional energy surface of a ChLC in external electric and magnetic fields is introduced as a function of spherical coordinates defining the director orientation. Reduced, two dimensional energy surface as a function of the first Fourier components of the spherical coordinates, presented in the same section provides an easy way to visualize the Fr\'{e}edericksz transition. An overview of the GNEB method for calculating MEPs on the multidimensional energy surface of ChLC is given in Sec.~\ref{sec3}. In Sec.~\ref{sec4}, the GNEB method is applied to a transition between P and D states of the ChLC and the effect of a thermal activation is estimated.

\section{Energy surface of a cholesteric liquid crystal}
\label{sec2}
\subsection{ Oseen-Frank model}
\label{sec2a}

A flat liquid crystal cell of thickness $L$ is considered. The $Z$-axis of the reference frame is chosen to be perpendicular to the cell plane.
 The system is assumed to be homogeneous in the $XY$-plane so that the director is a function of the $z$-coordinate only, ${\bf n}({\bf r}) = {\bf n} (z)$. The energy per unit area of the system is given by a sum of three terms:
\begin{equation}
\label{Ftot}
{\EuScript F}_{\mathrm{tot}}={\EuScript F}_{\mathrm e}+{\EuScript F}_{\mathrm f}+{\EuScript F}_{\mathrm{sf}},
\end{equation}
where each term is a functional of spherical coordinates $\theta (z)$ and $\phi(z)$ defining orientation of director ${\bf n} (z)$.
The first term in Eq.~(\ref{Ftot}) is associated with distortions of ChLC and can be written as~\cite{Valkov2013}:
\begin{equation}
\label{Frank}
\EuScript F_{\mathrm e}= 
\frac{1}{2} \int\limits_{0}^{L}
\left[{\EuScript A}(\theta)(\theta')^2 + {\EuScript B}(\theta)(\phi')^2-2{\EuScript C}(\theta)\phi'\right]\mathrm{d}z.
\end{equation}
Here 
prime denotes derivative with respect to $z$ and functions ${\EuScript A}(\theta)$, ${\EuScript B}(\theta)$, ${\EuScript C}(\theta)$ are 
defined as
\begin{align}
{\EuScript A}(\theta)&=K_{11}\sin^2\theta+K_{33}\cos^2\theta,\label{defA}\\
{\EuScript B}(\theta)&=\sin^2\theta(K_{22}\sin^2\theta+K_{33}\cos^2\theta),\label{defB}\\
{\EuScript C}(\theta)&=q_0 K_{22}\sin^2\theta,\label{defC}
\end{align}
where $K_ {ii}$
are Frank modules ($i = 1,2,3$) and $\pi/q_0$ is the helix pitch.

Second term in Eq.~\eqref{Ftot} represents the contribution from the external field~\cite{Valkov2013}:

\begin{equation}
\label{Ff}
{\EuScript F}_{\mathrm f}=
\begin{dcases}
-\frac{1}{2}\int_{0}^{L}\chi_a H^2\cos^2\theta\,\mathrm{d}z, & \text{for $H$-field},\\
-\frac{1}{8\pi}U^2\left(\int_{0}^{L}{\EuScript E}(\theta)\,\mathrm{d}z\right)^{-1}, &  \text{for $E$-field}.
\end{dcases}
\end{equation}
Here $H$ is the magnitude of external magnetic field, $\chi_a$ is the anisotropy of magnetic susceptibility defined as a difference between its longitudinal and transverse components, $U$ is the voltage applied at the boundaries of the ChLC cell and the function ${\EuScript E}(\theta)$ is defined as follows:
\begin{equation}\label{EuScriptE=}
 {\EuScript E}(\theta)=\left(\varepsilon_\perp+\vea\cos^2\theta\right)^{-1},
\end{equation}
where $\vea$ and $\varepsilon_\bot$ are the anisotropy and transverse component of dielectric permittivity, respectively. 

Third term in Eq.~\eqref{Ftot} is the surface energy per unit area, which
 is defined in terms of Rapini-Popular potential describing anchoring
with the cell boundaries~\cite{Rapini69}:

\begin{equation}
\label{rapini}
{\EuScript F}_{\mathrm{sf}}=
\frac{1}{2}\sum\limits_{s}\left[W_{\theta}^{s}\left(\theta-\theta_{0}^s\right)^2
+ W_{\phi}^{s}\left(\phi-\phi_0^s\right)^2\right].
\end{equation}
Index $s$ can take two values, $l$ for the lower boundary of the cell and $u$ for the upper boundary. $W_{\theta}^{s}>0$ and $W_{\phi}^s>0$ are the anchoring coefficients and the $\theta_{0}^s$ and $\phi_{0}^s$ are angles defining the easy directions for the director at the boundaries. If $W_{\theta}^s , W_{\phi}^s \gg {\EuScript F}_{\mathrm e}+{\EuScript F}_{\mathrm f}$, the director is fixed at the boundaries. This case corresponds to the rigid boundary conditions. Otherwise, the director may deviate from the easy directions at the boundaries (soft boundary conditions).

\subsection{Multidimensional energy surface}
\label{sec2c}
\begin{figure}[h]
\begin{center}
  \includegraphics[width=0.95\linewidth]{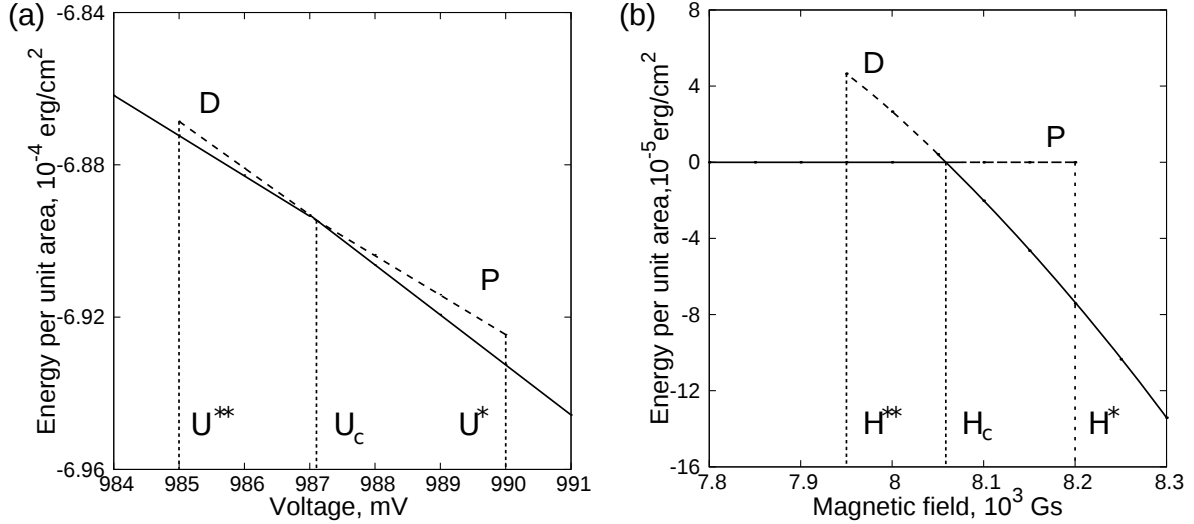}
\caption{Energy of the ChLC as a function of applied voltage (a) and magnetic field (b). Curves P and D correspond to planar and distorted states, respectively. Metastable states are shown with dashed curves.}
\label{fig_2}
\end{center}
\end{figure}

The energy surface of the ChLC can be introduced by applying the coarse-grained approximation to the system. The ChLC is divided into $N$ layers lying in the $XY$ plane and in each layer the director is assumed to be constant. 
Configuration of the system is then described by the set of spherical coordinates $\theta_i$ and $\phi_i$ defining the orientation of the director $\bf n_i$ in each element of the ChLC, $\bm{\Psi}\equiv\left(\theta_1,\phi_1,\theta_2,\phi_2,\ldots,\theta_N,\phi_N\right)$. 
The total energy as a function of $2N$ spherical coordinates, ${\EuScript F}_{\mathrm{tot}}={\EuScript F}_{\mathrm{tot}}(\bm{\Psi})$, can be obtained by applying Simpson's approximation to the integrals in Eqs.~\eqref{Frank} and \eqref{Ff}, where spatial derivatives are approximated using forward finite differences:
$
\psi_i^\prime \approx (\psi_{i+1}-\psi_i)/\Delta z,
$ 
 $\psi\equiv \theta,\phi$ and $\Delta z = L/(N-1)$. Function ${\EuScript F}_{\mathrm{tot}}(\bm{\Psi})$ defines a $2N$-dimensional energy surface where minima correspond to stable configurations. P state with director in plane parallel to cell surface and distorted (D) state, when it has non-zero out of plane projection, are such stable configurations. Depending on the parameters of ChLC, energy minima associated with P and D states can co-exist~\cite{Valkov2013}. This is illustrated by Fig.~\ref{fig_2}, where the energy of P and D states as a function of magnetic field and applied voltage is shown for a ChLC characterized by the following set of parameters:
$K_{11}=4.2\cdot10^{-7}$~dyn, $K_{22}=2.3\cdot10^{-7}$~dyn, $K_{33}=5.3\cdot10^{-7}$~dyn, $W_{\theta}^{l}=2.5\cdot10^{-3}$~{erg}/{cm$^2$}, $W_{\theta}^{u}=5.0\cdot10^{-4}$~{erg}/{cm$^2$};
$W_{\phi}^{l}=2.5\cdot10^{-4}$~{erg}/{cm$^2$},  $W_{\phi}^{u}=1.0\cdot10^{-4}$~{erg}/{cm$^2$};
$L=60$~$\mu$, $q_0=0.5$~{rad}/{$\mu$}; $\epsilon_\bot=7.2$, $\epsilon_a=9.0$; $ \theta_0^l=\theta_0^u=\frac{\pi}{2}$, $\phi_0^l=0$, $\phi_0^u=3$,  for system in electric field.  
$K_{11}=1.0\cdot10^{-6}$~dyn, $K_{22}=0.5\cdot10^{-6}$~dyn, $K_{33}=2.0\cdot10^{-6}$~dyn; 
$L=2$~$\mu$, $q_0=1.57$~{rad}/{$\mu$}; $\chi_a=10^{-7}$;  $ \theta_0^l=\theta_0^u=\frac{\pi}{2}$, $\phi_0^l=0$, $\phi_0^u=\pi$, for system in magnetic field with rigid boundary conditions. The discontinuous Fr\'{e}edericksz transition is expected in ChLC with this set of parameters~\cite{Valkov2013}.

At the zero voltage~(see Fig.~\ref{fig_2}(a)), only one stable state exists in the system which corresponds to a planar configuration of the director. The D configuration emerges at $U=U^{\ast\ast}=985$ mV as a metastable state. The energy of D state decreases faster with the voltage as compared to that of P state and at $U=U_c=987$ mV the energy levels of both states coincide. If the voltage further increases, the D state becomes energetically favorable. At  $U=U^\ast=990$ mV, the P state becomes unstable. ChLC in the external magnetic field shows similar behaviour and becomes bistable in the magnetic field range $H^{\ast\ast} <H <H^{\ast}$, where $H^{\ast\ast}=7950$ Gs,  $H_c=8060$ Gs,  $H^{\ast}=8200$ Gs (see Fig.~\ref{fig_2}(b)).

The ChLC initially prepared in the P state can be transferred to the D state by increasing the applied voltage (magnetic field). If the effect of thermal fluctuations is not taken into account, the transition from P to D occurs at $U=U^\ast$ ($H=H^\ast$) when the energy barrier separating the states vanishes. Inverse transition form D to P occurs at $U=U^{\ast\ast}$ ($H=H^{\ast\ast}$). However, temperature renormalizes transition fields: thermal fluctuations can induce spontaneous transitions even when the barrier is not zero. 
Specifically, transition field at a given temperature is the magnitude of the external field at which the time scale of thermally activated transitions from P (D) state to D (P) state, $\tau$, becomes
equal to the time scale of the experiment, $\tau_{exp}$.
The thermal lifetime $\tau$ can be estimated using the harmonic rate theory~\cite{coffey_01} which predicts the Arrhenius dependence on the temperature:
\begin{equation}
\tau = \tau_0 e^{\Delta E/k_BT},
\label{arrhen}
\end{equation}
where the pre-exponential factor $\tau_0$ is expected to weakly depend on the field and the energy barrier $\Delta E$ given by the energy difference between the local energy minimum and the first order saddle point (SP) is strongly field-dependent (see below). Eq.~\eqref{arrhen} is an implicit definition of the transition fields in ChLC at a finite temperature. 

Study of the effect of thermal fluctuations on the transitions in ChLC essentially becomes a problem of identifying the SPs on the energy surface. The first order SP is a stationary point on the energy surface which is a maximum with respect to one and only one degree of freedom, but a minimum with respect to the other degrees of freedom. One approach for locating SPs is based on finding minimum energy paths (MEPs) between given stable states, because maximum on the MEP is a SP on the energy surface. This approach is used here to study activation energy barriers for the transitions in ChLC.

Before we proceed with the analysis of MEPs and SPs on the multidimensional energy surface of a ChLC, we present a reduced, two-dimensional model of a ChLC with rigid boundary conditions in a magnetic field, where stable states, SPs and MEPs can be visualized easily giving a valuable insight into the mechanism of transitions in ChLC.

\subsection{Two-dimensional energy surface}
\label{sec2b}
Fourier components of the spherical coordinates $\theta(z)$, $\phi(z)$ can be used to define the energy surface of a ChLC. The dimensionality of the energy surface is then defined by the number of Fourier components taken into account in the analysis. In the simplest approximation, when only one Fourier harmonic is taken for $\theta(z)$ and for $\phi(z)$, the energy surface is two-dimensional and, therefore, can be visualized. In this case following functional form for the spherical coordinates may be used:
\begin{align}
\theta(z)&=\frac{\pi}{2} + c_\theta\cos\frac{\pi (z-L/2) }{L} \label{eq:theta}\\ 
\phi(z)&=q_0z + c_\phi\sin \frac{2\pi (z-L/2)}{L}.\label{eq:phi}
\end{align}

\begin{figure}[h]
\begin{center}
  \includegraphics[width=1.0\linewidth]{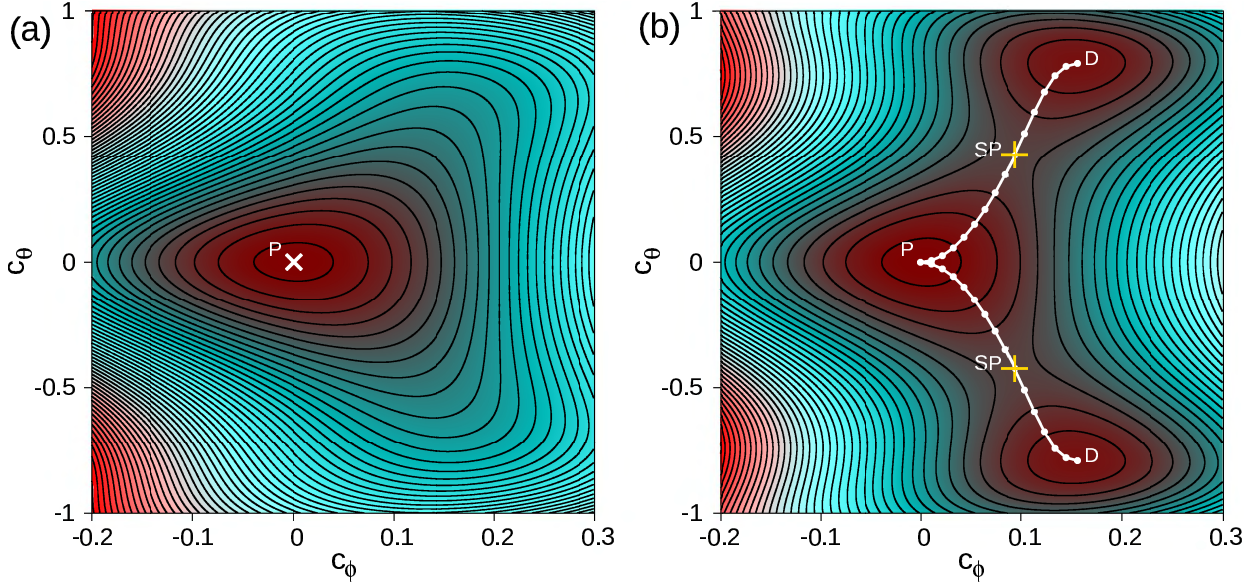}
\caption{Energy surface of the ChLC in magnetic field as function of the first non-zero Fourier components, $c_\theta$ and $c_\phi$, of the spherical coordinates $\theta(z)$ and $\phi(z)$ defining orientation of the director. (a) $H<H^{\ast\ast}$. (b) $H^{**} <H <H^{*}$. MEP is shown with a white curve while SPs are indicated with yellow crosses.}
\label{fig_1}
\end{center}
\end{figure}

A contour graph of the energy surface can be constructed by substituting $\theta(z)$ and $\phi(z)$ from Eqs.~\eqref{eq:theta} and~\eqref{eq:phi} into the expression for the energy of the system, Eq.~\eqref{Ftot}. The resulting enegy surface is obtained for two values of the magnetic field, $H = 7800$ Gs and $H = 8070$ Gs.
The structure of the energy surface depends on the magnitude of the external field. When $H<H^{\ast\ast}$ (Fig.~\ref{fig_1}(a)), there is only one minimum on the energy surface ($c_\theta=c_\phi=0$) which corresponds to P state. At larger fields ($H^{**} <H <H^{*}$), two equivalent stable D states appear in the system, while the P state is also present (see Fig.~\ref{fig_1}(b)). MEPs between the states pass through the SPs which define the energy barriers. 

Although the two-dimensional model of ChLC reveals the main characteristics of transitions between the stable states, the quantitative analysis of energy barriers requires calculations of stable states, MEPs and SPs for the full, multidimensional model. Locating the SPs on multidimensional energy surfaces is significantly more difficult than finding the minima. The difficulty arises from the need to minimize the energy with respect to all but one degree of freedom for which a maximization should be carried out. The problem is that it is not known a priori which degree of freedom should be treated differently. In the next section we briefly describe an efficient approach based on calculation of MEPs.

\section{Minimum energy paths}
\label{sec3}

An MEP between two minima is the path in the configuration space which lies lowermost on the energy surface. In the case of  Fr\'{e}edericksz transition in ChLC, following a MEP means rotating the director of each element of the ChLC in such a way that the energy is minimal with respect to all degrees
of freedom perpendicular to the path. A maximum along a MEP corresponds to a first order saddle point on the energy surface, and the highest maximum gives an estimate of the activation energy barrier. The MEP not only gives the position of SP, but also provides information about mechanism of the transition, as it represents the path of highest statistical weight. 

Special attention needs to be taken when calculating MEPs for the transitions in ChLC, because of the curvature of the configuration space arising from the constraint on the length of the director, $|\mathbf{n}_i| = 1$. Namely, configuration space of an ChLC divided into $N$ elements is a $2N$-dimensional Riemannian manifold, $\mathcal{R}$, corresponding to the direct product of $N$ 2-dimensional spheres:
\begin{equation}
\label{eq:conf_space}
\mathcal{R} = \prod_{i=1}^N S_i^2,
\end{equation}
where $S_i^2$ is a $2$-dimensional unit sphere associated with the director of the $i$-th element. Similar problem arises when studying transitions in magnetic systems, where a constraint is usually applied on the length of magnetic moments. 

Recently, the geodesic nudged elastic band (GNEB) method has been formulated to find MEPs in curved manifolds such as $\mathcal{R}$ and applied to transitions in magnetic systems~\cite{bessarab_2015}. Similar to the original nudged elastic band (NEB) method~\cite{NEBleri} widely used to study thermally induced atomic rearrangements, the GNEB 
method involves taking some initial guess of a path between the two minima,
and systematically bringing that to the nearest MEP. A path is
represented by a discrete chain of states, or 'images', of the system, where the first and the last image are placed at the energy minima corresponding to the stable configurations. In order to distribute the images evenly along the path, springs are introduced between adjacent images. At each image, a local tangent to the path needs to be estimated 
and the force guiding the images towards the nearest MEP is defined 
as the sum of the transverse component of the negative energy gradient plus the component of the spring force along the tangent.   
The position of intermediate images is then adjusted so as to zero the GNEB force. 

An important aspect of the method is that both the GNEB force and the path tangent are defined in the local tangent space of the $\mathcal{R}$ manifold, which is needed to satisfy the constraint on the length of the director and to properly decouple the perpendicular component of the energy gradient from the spring force~\cite{bessarab_2015}.

A more detailed description of the GNEB method applied to ChLC in a planar cell is as follows. A chain of $Q$ images is constructed, $\left[\bm{\Psi}^1,\bm{\Psi}^2,\ldots,\bm{\Psi}^Q\right]$, 
where the endpoints are fixed and given by the 
local minima corresponding to P and D configurations in the ChLC, while the $Q-2$ intermediate images 
$\bm{\Psi}^\nu=\left(\theta_1^\nu,\phi_1^\nu,\ldots,\theta_N^\nu,\phi_N^\nu\right)$, $\nu=2,\ldots,Q-1$, give a discrete representation of a path. The position of the intermediate images is adjusted in order to converge on the MEP. This is accomplished by systematically displacing the images in the direction defined by the GNEB force acting on them so as to zero this force. 
The GNEB forces $\bm{F}^\nu_{GNEB}$ guiding the images towards the MEP are defined as follows:
\begin{equation}
\label{eq:forceGNEB}
\bm{F}^\nu_{GNEB} = ( - \nabla E\left(\bm{\Psi}^\nu\right)|_\perp + \bm{F}^\nu_s|_\parallel)_\mathcal{T}.
\end{equation}
Here the subscript $\mathcal{T}$ denotes projection of a vector on the local tangent space of $\mathcal{R}$. The perpendicular component of the energy gradient is obtained by subtracting out the parallel component
\begin{equation}
\label{eq:Gperp}
\nabla E\left(\bm{\Psi}^\nu\right)|_\perp = \nabla E\left(\bm{\Psi}^\nu\right)-\left(\nabla E\left(\bm{\Psi}^\nu\right)\cdot\hat{\bm{\tau}}^\nu_\mathcal{T}\right)\hat{\bm{\tau}}^\nu_\mathcal{T},
\end{equation}
where the unit tangent to the path, $\hat{\bm{\tau}}^\nu_\mathcal{T}$, lies in the tangent space, which is indicated by the subscript $\mathcal{T}$.
The parallel component of the spring force is evaluated as
\begin{equation}
\label{eq:GFsparallel}
\bm{F}^\nu_s|_\parallel = \kappa\left[L\left(\bm{\Psi}^{\nu+1},\bm{\Psi}^{\nu}\right)-L\left(\bm{\Psi}^{\nu},\bm{\Psi}^{\nu-1}\right)\right]\hat{\bm{\tau}}^\nu_\mathcal{T}.
\end{equation}
Here, $L\left(\bm{\Psi}^{\nu+1},\bm{\Psi}^{\nu}\right)$ and $L\left(\bm{\Psi}^{\nu},\bm{\Psi}^{\nu-1}\right)$ are geodesic distances between images $\nu+1$, $\nu$ and $\nu$, $\nu-1$, respectively, and $\kappa$ is a spring constant. Since the spring force is decoupled from the perpendicular component of the energy gradient, the value of the spring constant is not critical and, in fact, can be varied over several orders of magnitude without affecting the calculation results~\cite{bessarab_2015}.

Some minimization method needs to be used in connection with the GNEB method so as to zero the forces $\bm{F}^\nu_{GNEB}$. We used the velocity projection optimization (VPO) algorithm based on a fictitious equation of motion
of a point mass on a curved manifold $\mathcal{R}$ where the velocity is damped by
including only the component in the direction of the force~\cite{bessarab_2015}. Once convergence has been reached, the images lie on the MEP where
the energy gradient $\nabla E\left(\bm{\Psi}^\nu\right)|_\mathcal{T}$ can only have a component in the direction of the path. The activation energy and the SP configuration can then be derived from the maximum along the MEP.

\section{Results}
\label{sec4}

\begin{figure}[h]
\includegraphics[]{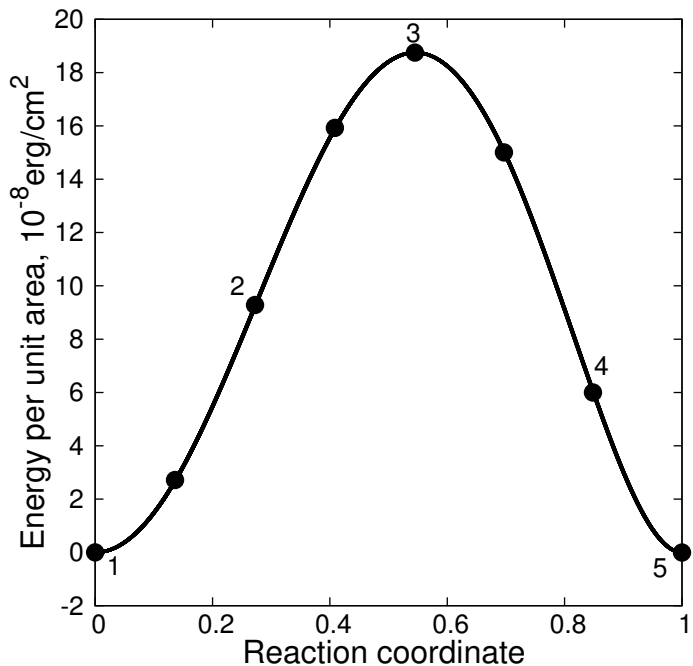}
\caption{Energy per unit area along the MEP at $U=U_c=986.84$ mV. The filled circles correspond to the images of the system used in the GNEB calculation. The reaction coordinate
is defined as the displacement along the path normalized by its total length.
}
\label{fig3}
\end{figure}

The GNEB method was applied to the ChLC in external electric field, for which parameter values listed in Sec.~\ref{sec2}B were used. For each value of the applied voltage, both P and D state were found by minimizing the total energy of the system and the MEP between them was identified. Fig.~\ref{fig3} shows the energy change along the MEP for the transition between P and D states in the ChLC, where the magnitude of applied voltage was chosen to be $U=U_c=986.84$ mV, at which the energy levels of P and D states coincide. Maximum along the MEP gives the energy barrier for the transition. Position of the maximum along the MEP~Fig.~(\ref{fig3}) was found using Climbing Image GNEB~\cite{bessarab_2015}.

\begin{figure}[h]
\centering
  \includegraphics[width=1.0\linewidth]{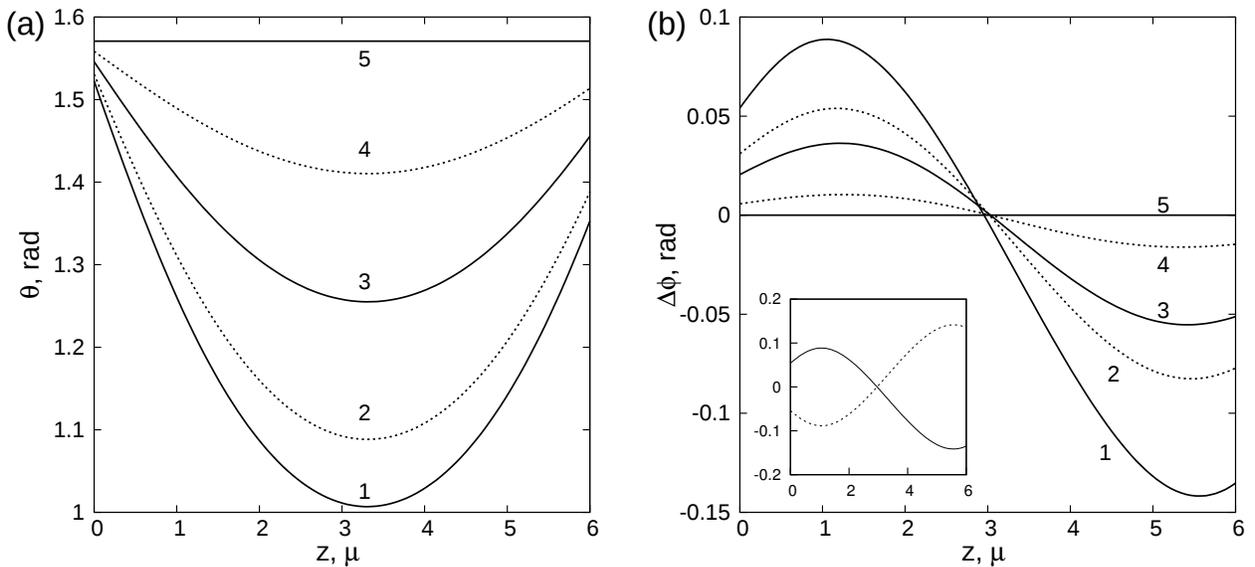}
\caption{Profiles of the polar angle (a) and deviation of the azimuthal angle from a straight line, $\phi(z)=q_0z$, (b) for the images along the MEP. Labeling of the curves correspond to that of the points in Fig.~\ref{fig3}. The inset (b): $\Delta\phi (z)$ for the opposite twist of the ChLC .}
\label{fig4}
\end{figure}

While the images at the ends of the MEP correspond to stable configurations in the ChLC (P and D states),  intermediate images provide information about changes in the system during the transition between the states. Fig.~\ref{fig4} demonstrates intermediate configurations of the ChLC during the transition from P to D state for $U=U_c$. Each configuration is shown as a profile of spherical coordinates, $\theta (z)$ (Fig.~\ref{fig4}(a)) and $\phi(z)$ (Fig.~\ref{fig4}(b)). The straight line in Fig.~\ref{fig4}(a) $\theta=\pi/2$ is the P state, and the curve with the largest deviation from that line is the D state. Curve 3 corresponds to the configuration of the ChLC at the saddle point. The minimum of $\theta (z)$ does not change its position on the $z$-axis along the MEP. For the rigid boundary conditions, this minimum is in the middle of the cell, but it is shifted towards a boundary with a smaller anchoring coefficient for the soft boundary conditions.
For each image along the MEP, the azimuthal angle $\phi (z)$ demonstrates small deviation from the straight line $\phi (z)= q_0z$. Figure~\ref{fig4}(b) shows the magnitude of this deviation, $\Delta\phi (z)$, as a function of $z$ for several images along the MEP between P and D states. $\Delta\phi (z)$ is antisymmetric with respect to the center of the cell for the rigid boundary conditions. However, the symmetry is broken in the case of soft boundary conditions.

The chirality of the system is defined by the parameter $q_0$. If $q_0>0$, then $\phi(z)$ changes clockwise. If $q_0<0$, the chirality becomes anticlockwise. The inset in Fig.~\ref{fig4}(b) demonstrates the profile $\Delta\phi (z)$ for the opposite twist of the ChLC director, i.e. after replacement $q_0 \longrightarrow -q_0$.

\begin{figure}[h]
\includegraphics[width=1.0\linewidth]{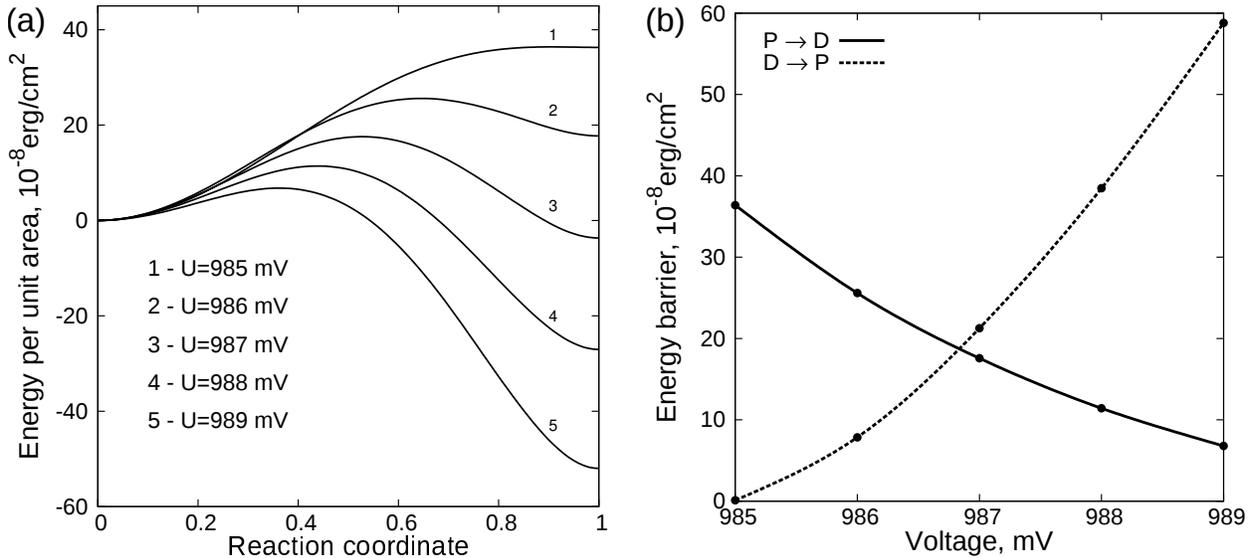}
\caption{(a) MEPs for the transition between P and D states at different applied voltages. The reaction coordinate
is defined as the displacement along the path normalized by its total length. (b) Energy barrier for transition from the P to D state (solid line) and for the inverse transition (dashed line) as a function of applied voltage.}
\label{fig5}
\end{figure}

Changes in external electric field lead to the changes in the shape of the energy surface and, therefore, MEPs between states as well as the corresponding energy barriers.
Fig.~\ref{fig5}(a) shows the MEPs for five various magnitudes of applied voltage. At $U=U^{\ast\ast}$ (curve 1) the energy along the MEP is completely flat at one of the ends of the path, which is a signature of the emergence of D state. The D state energy minimum becomes more pronounced as the voltage increases, while the P state minimum becomes shallower (curve 2). Therefore, there is a threshold voltage, $U=U_c$, at which the energy levels of both configurations coincide (see Fig.~\ref{fig3}). If the voltage is further increased, P configuration becomes a metastable state (curves 3, 4 and 5).
Although the P state is not a ground state of the system in the voltage range between $U_c$ and $U^\ast$, there is still a finite barrier separating this state from D state. If the system is initially prepared in P state and effect of temperature is not included, this barrier prevents the system from passing to D state even if voltage is close to $U^\ast$ and transition occurs only if the barrier vanishes, i.e. the voltage reaches the value of $U^\ast$. However, thermal fluctuations can be sufficient to induce over-the-barrier transitions on the laboratory time scale, which renormalizes the transition voltage. 

Barrier for the P$\rightarrow$D transition monotonically decreases to zero with the applied voltage, while the barrier for the inverse transition gets larger as the voltage is increased. Two curves intersect at $U=U_c$ Fig.~\ref{fig5}(b).

\begin{figure}[h]
\centering
  \includegraphics[width=1.0\linewidth]{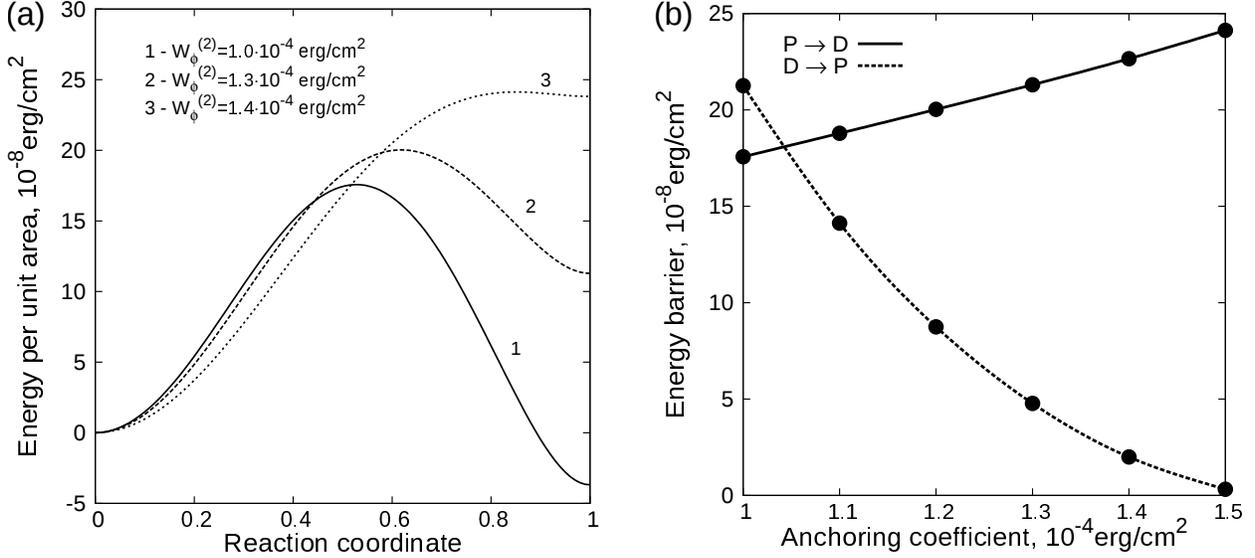}
\caption{(a) The MEPs for transition from the P to the D state for different anchoring coefficients $W_\phi^{u}$ at the upper boundary. The reaction coordinate is defined as the displacement along the path normalized by its total length. (b) Energy barrier for transition from the P to D state (solid line) and for the inverse transition (dashed line) as a function of $W_\phi^{u}$. Voltage is taken to be  $U=987$mV.}
\label{fig6}
\end{figure}

A variation of anchoring coefficients also changes the energy surface. Dependence of MEP on the anchoring coefficient $W_\phi^{u}$ while other anchoring coefficients are kept fixed is shown in~figure~\ref{fig6}(a). 
The energy of the D state increases monotonously compared to that of the P state as the anchoring coefficient at the upper boundary gets larger. Thus, variation of the anchoring coefficient has similar effect as variation of voltage: there is a threshold value of $W_\phi^{u}$ at which the energy levels of both states are the same. Further increase in $W_\phi^{u}$ makes D state metastable first (curve 2) and then unstable (curve 3). Barriers for between P and D state also strongly depend on the anchoring coefficient.

Similar behavior takes place in an external magnetic field. Thus, the voltage, the magnetic field and anchoring coefficients can be used for tuning the energy barrier separated two states. 

These results can now be used to estimate how characteristics of the Fr\'{e}edericksz transition change with temperature. Usually such influence is explained in terms of temperature dependent adjustment of the Frank modules or other parameters of a liquid crystal. Indeed, the variation of Frank modules modifies the energy surface and, therefore, may change the transition field. However, quantitative assessment of thermally activated transitions between states in liquid crystal can explain the effective renormalization of parameters of the Fr\'{e}edericksz transition. 

In particular, the temperature dependence of the transition voltage $U^\ast$ can be explained as follows. The system will remain in P state until the applied voltage has lowered the energy barrier sufficiently and, thereby, decreased the lifetime of P state sufficiently for the transition to D state to occur on the laboratory time scale.  According to Eq.~(\ref{arrhen}), the lifetime is mostly defined by the energy barrier, $\Delta E = \Delta E(U)$ which is strongly voltage-dependent (see Fig.~\ref{fig5}(a)). Assuming constant pre-exponential factor, the change in $U^\ast$ can be predicted from the the following equation that can be obtained from the Arrhenius fromula (see Eq.~(\ref{arrhen}))
\begin{equation}
\frac{\Delta E(U^\ast_1)}{\Delta E(U^\ast_2)}=\frac{T_1}{T_2},
\label{eq:barr}
\end{equation}
an implicit expression showing how $U^\ast$ changes with temperature. Eq.~(\ref{eq:barr}) predicts a drop of $0.5$ mV for a $U^\ast$ as temperature is raised from $300$ K to $360$ K.(see Fig.~\ref{fig5}(b))

In summary, we have introduced the multidimensional energy surface of ChLC in a planar cell as a function of spherical coordinates which determine the orientation of director profile across the cell. In a certain range of external electric (magnetic) field energy surface contain two local minima corresponding to P and D states of ChLC. Transition between these states is the discontinuous Fr\'{e}edericksz  effect. MEP between P and D phases gives the energy barrier which needs to be overcome for such transition. The height of barrier and the shape of energy surface strongly depend on the applied field and boundary conditions. In particular, the energy barrier can be lowered so that thermal fluctuations become sufficient to stimulate the Fr\'{e}edericksz  transition. It gives an additional contribution to temperature dependence of characteristics of Fr\'{e}edericksz transition.

\section{Acknowledgements}
This work was supported by Saint Petersburg State University (Grant No. 11.37.145.2014), the Russian Foundation of Basic Research (Grant No. 14-02-00102), Nordic-Russian Training Network for Magnetic Nanotechnology(NCM-RU10121). PB gratefully acknowledges support from the G\"{o}ran Gustafsson Foundation.

\end{document}